\documentclass[12pt,a4paper]{article}
\usepackage[utf8]{inputenc}
\usepackage[english]{babel}
\usepackage{amsmath}
\usepackage{amsfonts}
\usepackage{amssymb}
\usepackage{graphicx}
\usepackage{graphics}
\usepackage{epsfig}
\begin{document}

\title{
\begin{flushright}
{\small INR-TH-2014-008}
\end{flushright}
    H\'{e}non-Heiles potential as a bridge between nontopological solitons
    of different types
  }
\author{
 E.\,Ya.\;Nugaev\thanks{{\bf e-mail}:
emin@ms2.inr.ac.ru}\\
{\small{\em
Institute for Nuclear Research of the Russian Academy
of Sciences,}}\\
{\small{\em 60th October Anniversary prospect 7a, Moscow 117312,
Russia
}}
}
\maketitle

\begin{abstract}
We apply the Hubbard-Stratanovich transformation to the Lagrangian for nontopological
solitons of the Coleman type in a two-dimensional theory. The resulted theory with
an extra real scalar field can be supplemented with a cubic term to obtain
a model with exact analytical solution. 
\end{abstract}

\section{Introduction}

Internal symmetry may provide a mechanism of stabilization of extended objects 
in a scalar field theory with an appropriate self-interaction~\cite{Rosen,Coleman:1985ki}. Constructions with extra
scalar fields are also possible~\cite{Friedberg:1976me} depending on the 
existence of a second symmetry which is spontaneously broken\footnote{See also references on earlier works in~\cite{Friedberg:1976me}.}. Following~\cite{Coleman:1985ki} we will say that the latter theories and theories with only one scalar field  are of different genus. In four space-time dimensions
such spherically symmetric solutions are usually referred to as Q-balls of the Coleman and
Friedberg-Lee-Sirlin types respectively.

In this paper we will study nontopological solitons in two-dimensional theories providing analytical solutions. We will start from the theory of one complex scalar field with self-interaction. Consideration of this self-interaction as an effective interaction due to the massive field leads to the theory with an extra
scalar field similar to the Friedberg-Lee-Sirlin case. We show that the
theory of a single complex scalar field can be completed in a simple way to obtain an explicit solution for
the nontopological soliton.

\section{Nontopological soliton of the Coleman type}
Let us start with the simplest model of a scalar field
in two dimensions. The $U(1)$-invariant Lagrangian density is
\begin{equation}
{\cal L}\equiv T-V= \partial_\mu\phi^*\partial^\mu\phi-m^2(\phi^*\phi)+\frac{\lambda}{2}(\phi^*\phi)^2
\label{phi4}
\end{equation}
with positive\footnote{The potential $V$ is unbounded from below. However one 
can add positive term $\epsilon(\phi^*\phi)^3$ with $\epsilon\ll \lambda$ to cure the model for large values
of the field.} $\lambda$. Decay of the vacuum $\phi=0$ is exponentially suppressed in the
case of $\lambda/m^2\ll 1$, thus we can use density (\ref{phi4}) for the effective
description.

Q-balls in this model were examined in \cite{Belova:1994vd}
and the earlier works on soliton interaction. For the usual anzatz
\[
\phi(t,z)={\rm e}^{{\rm i}\omega t}f(z)
\]
one can obtain a solution of the form
\begin{equation}
f=\sqrt{\frac{2(m^2-\omega^2)}{\lambda}}\frac{1}{\cosh(\sqrt{m^2-\omega^2}z)}.
\label{sol_phi4}
\end{equation}

\begin{figure}[!hc]
\includegraphics[width=4in]{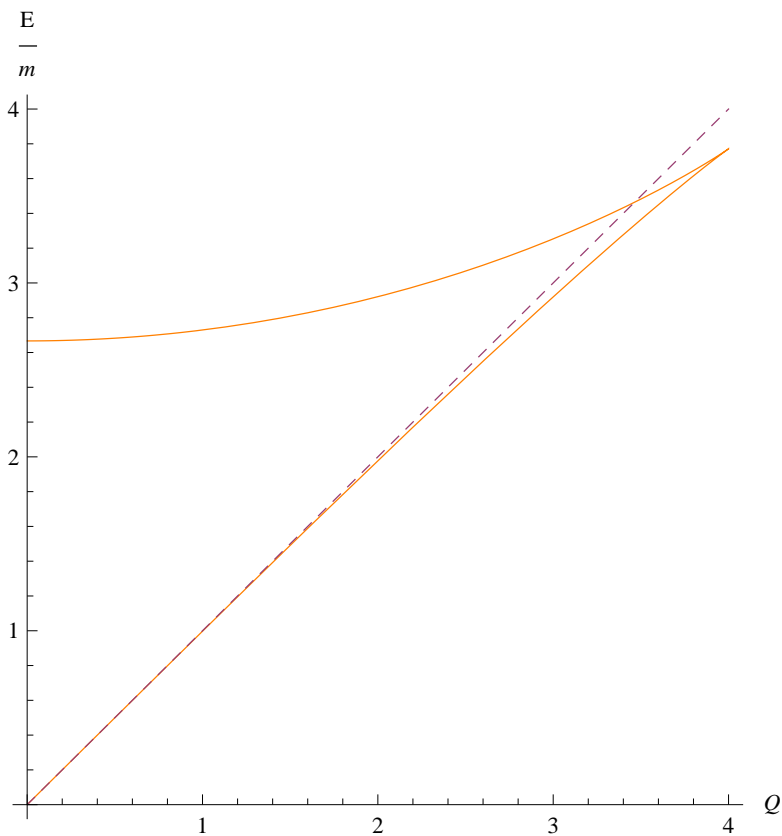}
\caption{The $E(Q)$ dependence for $\lambda=m^2$ (solid line). Dashed line 
corresponds to free particles with $E=mQ$.
} \label{EV4}
\end{figure}

For $\omega=0$ it corresponds to the bounce for the tunnelling problem\footnote{This feature can be used as a starting point for the demonstration of the existence of continuius family of solutions~\cite{Rajaraman:1975qr}.}. Energy $E$ and charge $Q$ are 
finite for all $\omega \in [0,m]$:
\begin{equation}
E=\frac{8}{\lambda}\sqrt{m^2-\omega^2}\left(m^2-\frac{2}{3}(m^2-\omega^2)\right),
\label{Ephi4}
\end{equation}
\begin{equation}
Q=\frac{8\omega}{\lambda}\sqrt{m^2-\omega^2}.
\label{Qphi4}
\end{equation}
On Fig.\ref{EV4} we plot the $E(Q)$ dependence for $\lambda=m^2$. To obtain this
dependence for
arbitrary $\lambda$ one can use the following scaling to dimensionless parameters:
\begin{equation}
\omega\to \frac{\omega}{m},\qquad
E\to\frac{E}{m}\frac{\lambda}{m^2}\qquad
Q\to Q\frac{\lambda}{m^2}.
\label{scaling_phi4}
\end{equation}
It is useful to introduce an analogue of the grand potential
\begin{equation}
\Omega(\omega)\equiv E(Q(\omega))-\omega Q(\omega)=\frac{8}{3\lambda}\left(\sqrt{m^2-\omega^2}\right)^3.
\label{omega_phi4}
\end{equation}
From obvious relation $Q=-d\Omega/d\omega$ one can obtain the equality $dE/dQ=\omega$,
which may be used to verify our result.
This relation is also useful for applying the classical stability 
criterion $d^2E/dQ^2<0$ (or, using (\ref{omega_phi4}), $\partial^2\Omega/\partial\omega^2>0$)
of \cite{LeePang}:
\[
\frac{d^2E}{dQ^2}=\frac{d\omega}{dQ}=\frac{8}{\lambda}\sqrt{m^2-\omega^2}\frac{1}{m^2-2\omega^2}.
\]
It means that the solution is stable classically for $\omega/m\in(1/\sqrt2,1)$. This range of parameters corresponds to the lower branch of Fig.\ref{EV4} with $Q_{min}=0$ and $Q_{max}=4m^2/\lambda$. Thus, for the small coupling regime $m^2/\lambda\gg 1$ one
can consider nontopological solitons with large $Q$ as quasiclassical objects.

\section{Analytical solution with extra field}

Interaction in~(\ref{phi4}) can be considered as an effective interaction due to
an extra very massive real scalar field $\chi$ \cite{Hubbard-Stratanovich}.
Let us consider the Lagrangian
\[
\partial_\mu\phi^*\partial^\mu\phi-m^2(\phi^*\phi)+\frac{1}{2}
\partial_\mu\chi\partial^\mu\chi-\frac{M^2}{2}\chi^2+g\chi(\phi^*\phi).
\]
In the limit $M^2\gg m^2$ the heavy field $\chi$ can be integrated out to obtain
an effective low-energy theory of a single field $\phi$. The tree-level result is equivalent to (\ref{phi4}) with 
self-interaction coupling constant
$\lambda_{eff}=\frac{g^2}{M^2}$.

One can find a solution in the system of coupled fields using the perturbation theory.
However, note that an exact analytical solution exists in the following model
\begin{equation}
{\cal L} = \partial_\mu\phi^*\partial^\mu\phi-m^2(\phi^*\phi)+\frac{1}{2}
\partial_\mu\chi\partial^\mu\chi-\frac{M^2}{2}\chi^2+g\chi(\phi^*\phi)
+g\chi^3.
\label{Henon-Heiles}
\end{equation}
Here the coupling constant $g$ is arbitrary, but the cubic interaction is tuned.
The solution for a nontopological soliton looks like
\begin{equation}
\phi={\rm e}^{{\rm i}\omega t}\frac{A}{{\rm cosh} (\sqrt{m^2-\omega^2}z)}, \qquad \chi=\frac{B}{({\rm cosh} (\sqrt{m^2-\omega^2}z))^2},
\label{solutionHH}
\end{equation}
where 
\begin{equation}
\begin{array}{c}
A=\frac{\sqrt{m^2-\omega^2}}{g}\sqrt{2(M^2-4(m^2-\omega^2))},\\
B=\frac{2(m^2-\omega^2)}{g}.
\label{const}
\end{array}
\end{equation}
The charge of this configuration
\begin{equation}
Q=\frac{8\omega \sqrt{m^2-\omega^2}}{g^2}(M^2-4(m^2-\omega^2))
\label{QHH}
\end{equation}
and the energy
\begin{equation}
E=\frac{8\sqrt{m^2-\omega^2}}{g^2}\left(\frac{2}{3}\omega^2M^2+\frac{1}{3}m^2M^2-2(m^4-\omega^4)+\frac{6}{5}(m^2-\omega^2)^2\right).
\label{EHH}
\end{equation}
also satisfy the constraint\footnote{This relation also holds for solitons
with arbitrary $V(\chi)$ in the Lagrangian (\ref{Henon-Heiles}).} $dE/dQ=\omega$. The Compton length of our solution is small compared to
its characteristic size for the intermediate values of $\omega/m$ for
$g/M^2\ll 1$.

The reason of integrability of equations of motion lies in the formal correspondence
to the problem of classical mechanics with the H\'{e}non-Heiles potential \cite{HHoriginal} (see also \cite{HH} for an explicit integral of motion for our choice of the cubic interaction).

\begin{figure}[h!]
\begin{minipage}[h]{0.49\linewidth}
\center{\includegraphics[width=1.0\linewidth]{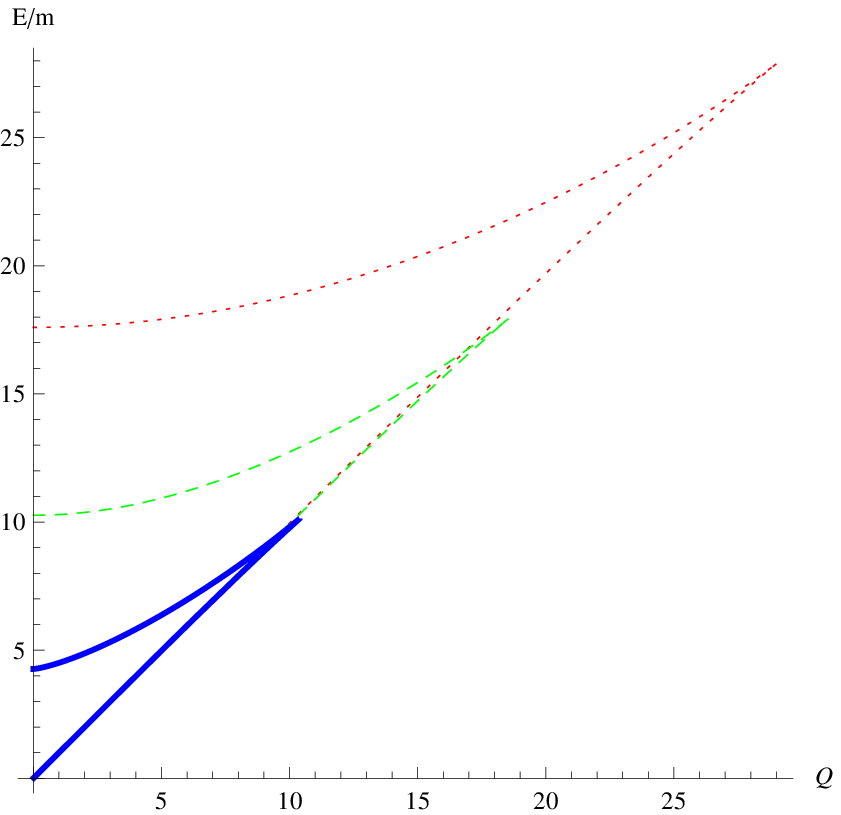}\\a}
\end{minipage}
\begin{minipage}[h]{0.49\linewidth}
\center{\includegraphics[width=1.0\linewidth]{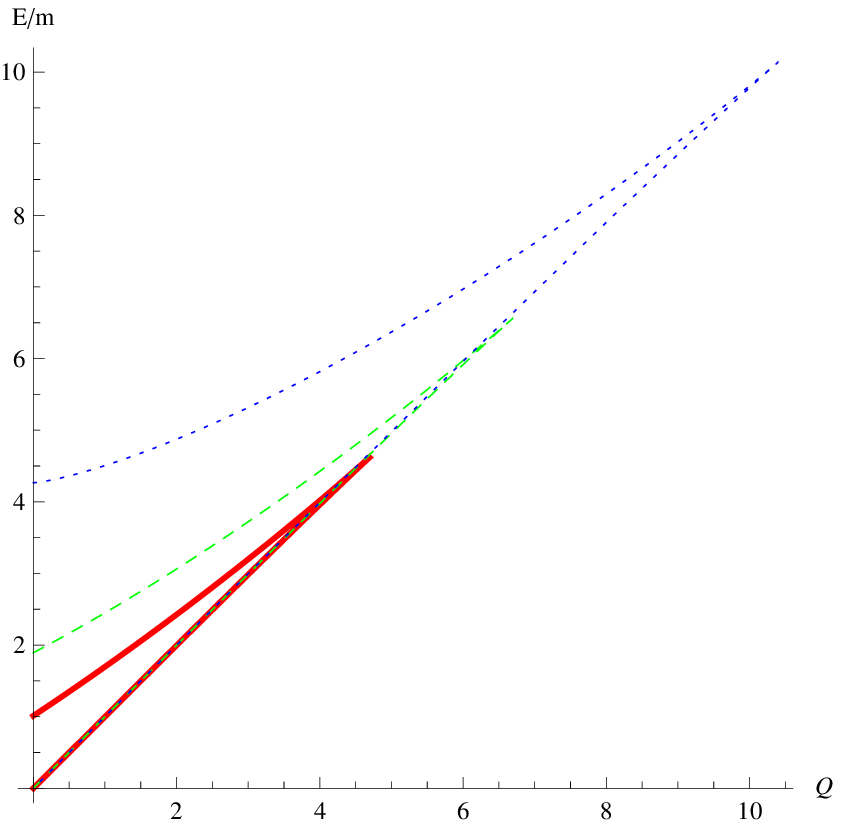}\\b}
\end{minipage}
\caption{The $E(Q)$ plots for $M/m\geq 2$, $g=m^2$ (left plot, $M/m=2$ -- thick line,
$M/m=2.5$ -- dashed line, $M/m=3$ -- dotted line) and $m/M\leq 2$, $g=m^2$ (right plot, $M/m=1.5$ -- thick line,
$M/m=1.7$ -- dashed line, $M/m=2$ -- dotted line). } 
\label{E(Q)HH}
\end{figure}

One can see from (\ref{const}) that the range of allowed values is 
$\omega/m\in[0,1]$ for $M/m\geq2$ and $\omega/m\in[\sqrt{1-(M/(2m))^2},1]$ for
$M/m\leq2$. From Fig.\ref{E(Q)HH} one can see that the $E(Q)$ dependence
reproduces
the shape of Fig.\ref{EV4} even for the value $m/M=0.33$ which is not very small. 
One can also obtain the values of the charge and the energy for the coupling constant $g\neq m^2$
using an analogue of reparametrization (\ref{scaling_phi4}).

\section{Conclusion}

New analytical solution for nontopological soliton with an extra scalar field is obtained. The integrability of equations of motion is determined by the formal
correspondence to the problem of classical mechanics with the H\'{e}non-Heiles symmetry.
There is no additional discrete symmetry in (\ref{Henon-Heiles}) which can be
spontaneously broken. In this sense our soliton is not of the Friedberg-Lee-Sirlin type. 
However, the absence of self-interaction for the field $\phi$ in (\ref{Henon-Heiles})
is crucial, because the existence of solution is determined by the extra field $\chi$. 
We examined the limit $m/M\to 0$ in which the theory with the simplest self-interaction
is reproduced in the tree-level approximation.
A remarkable feature of our model that is its renormalizability 
in theories with extra dimensions, i.e. cubic interactions
in (\ref{Henon-Heiles}) are irrelevant operators in theories with more than  three spatial dimension.

The author is indebted to M.~N.~Smolyakov and S.~V.~Troitsky for reading the text and providing comments that improved the manuscript, and M.~V.~Libanov and S.~Yu.~Vernov for helpful discussions and correspondence.
This work was supported by grant NS-2835.2014.2 of the President
of Russian Federation and by RFBR grant 14-02-31384.

\end{document}